%% file: sample-manuscript.tex
\documentclass[sigconf]{acmart}

\usepackage[draft=true]{minted}
\definecolor{bg}{rgb}{0.95,0.95,0.95}

\AtBeginDocument{%
  \providecommand\BibTeX{{%
    \normalfont B\kern-0.5em{\scshape i\kern-0.25em b}\kern-0.8em\TeX}}}

\copyrightyear{2024}
\acmYear{2024}
\setcopyright{acmlicensed}\acmConference[Q-SE 2024]{2024 ACM/IEEE International Workshop on Quantum Software Engineering }{April 16, 2024}{Lisbon, Portugal}
\acmBooktitle{2024 ACM/IEEE International Workshop on Quantum Software Engineering (Q-SE 2024), April 16, 2024, Lisbon, Portugal}
\acmDOI{10.1145/3643667.3648225}
\acmISBN{979-8-4007-0570-0/24/04}

\begin{document}

%%
%% The "title" command has an optional parameter,
%% allowing the author to define a "short title" to be used in page headers.
\title{Quantum types: going beyond qubits and quantum gates}

%%
%% The "author" command and its associated commands are used to define
%% the authors and their affiliations.
%% Of note is the shared affiliation of the first two authors, and the
%% "authornote" and "authornotemark" commands
%% used to denote shared contribution to the research.
\author{Tam\'as Varga}
\orcid{0000-0003-0479-9481}
\affiliation{%
  \institution{Constructor Institute Schaffhausen}
  \streetaddress{Rheinweg 9}
  \city{Schaffhausen}
  \country{Switzerland}
  \postcode{8200}
}
\email{tamas.varga@constructor.org}

\author{Yaiza Aragon\'es-Soria}\orcid{0000-0002-8880-8957}
\affiliation{%
  \institution{Constructor Institute Schaffhausen}
  \streetaddress{Rheinweg 9}
  \city{Schaffhausen}
  \country{Switzerland}
  \postcode{8200}
}
\email{yaiza.aragonessoria@constructor.org}

\author{Manuel Oriol}\orcid{0000-0003-4069-7626}
\affiliation{%
  \institution{Constructor Institute Schaffhausen}
  \streetaddress{Rheinweg 9}
  \city{Schaffhausen}
  \country{Switzerland}
  \postcode{8200}
}
\affiliation{%
  \institution{Constructor University Bremen}
  \streetaddress{Campus Ring 1}
  \city{Bremen}
  \country{Germany}
  \postcode{28759}
}
\email{mo@constructor.org}

%%
%% By default, the full list of authors will be used in the page
%% headers. Often, this list is too long, and will overlap
%% other information printed in the page headers. This command allows
%% the author to define a more concise list
%% of authors' names for this purpose.
\renewcommand{\shortauthors}{Varga, Aragon\'es-Soria and Oriol}

%%
%% The abstract is a short summary of the work to be presented in the
%% article.
\begin{abstract}
Quantum computing is a growing field with significant potential applications. 
Learning how to code quantum programs means understanding how qubits work and learning to use quantum gates. 
This is analogous to creating classical algorithms using logic gates and bits.
Even after learning all concepts, it is difficult to create new algorithms, which hinders the acceptance of quantum programming by most developers.

This article outlines the need for higher-level abstractions and proposes some of them in a developer-friendly programming language called Rhyme.
The new quantum types are extensions of classical types, including bits, integers, floats, characters, arrays, and strings.
We show how to use such types with code snippets.
\end{abstract}

%%
%% The code below is generated by the tool at http://dl.acm.org/ccs.cfm.
%% Please copy and paste the code instead of the example below.
%%


%%
%% Keywords. The author(s) should pick words that accurately describe
%% the work being presented. Separate the keywords with commas.

\received{7 December 2023}
%\received[revised]{XX XX 2024}
\received[accepted]{11 January 2024}

%%
%% This command processes the author and affiliation and title
%% information and builds the first part of the formatted document.
\maketitle

\section{Introduction}
Why are there so few algorithms for quantum computing? 
In our opinion, the adoption of quantum computing is hindered by the lack of high-level quantum abstractions to code and reason about problems.
Both current abstractions and tools are devised for the qubit level~\cite{Heim2020}.
For example, platforms such as Qiskit~\cite{Cross18} provide the means for developers to create quantum circuits using Python.
On the conceptual level, this is similar to coding with bits and logic gates.
Despite the limitations of the current hardware, it is already possible to create the abstractions that allow regular developers to create their own algorithms. 

This article proposes Rhyme, a high-level quantum programming language, that makes quantum programming accessible to software engineers by providing abstractions to develop quantum software in an intuitive manner.
We believe this will encourage developers to join the quantum software engineering community, and lead to new quantum applications and algorithms faster than using gate-level languages.

The following sections informally describe the proposed language via simple code snippets, highlighting the main features which differentiate it from the state-of-the-art. 
For the first version of the language, the compiler should generate code following OpenQASM 2.0~\cite{Cross17}, a machine-independent quantum assembly language for gate-based quantum computation. 
Several major quantum computing platforms have support for OpenQASM, including IBM Quantum~\cite{IBMQuantum}, Azure Quantum~\cite{AzureQuantum}, and Amazon Braket~\cite{AmazonBraket}.

Section~\ref{s:data_types} explains data types.
Section~\ref{s:computation} presents basic quantum-related operations.
Section~\ref{s:eval} describes some properties and shows how to code the Grover algorithm.
Section~\ref{s:rw} evaluates related work.
Section~\ref{s:conc} concludes and presents future work.

\section{Data types}\label{s:data_types}
The main intuition is to extend classical data types to quantum data types.
The basis states of a quantum data type are then equal to the possible values of the corresponding classical data type. 
Programmers can express superposition states by mapping probability amplitudes to classical data values and further manipulate such quantum data by re-mapping the amplitudes. 
Such (re-)mapping of the amplitudes can be achieved by writing classical functions using familiar high-level classical data types.

\textbf{Classical types.} Supported classical types are shown in the code example below. 
Arrays of classical bits are allowed. 
The bit-lengths of numeric types and the maximum character-length of strings are compiler parameters.

\vspace{4pt}
\begin{minted}[bgcolor=bg,frame=single,framesep=2mm]{text}
bit b = 0;
bit[] bArr; // not yet initialized
int i = 15;
float f = cos(pi / 4); // common math is allowed
complex z = 0.6 + 0.8i;
char c = 'A'; // any ASCII character
string s = "Hello";
bool t = true;
ref r = &i; // holds the address of i
\end{minted}
\vspace{4pt}

\textbf{Quantum types.} The quantum types are extensions of the classical types.
On a qubit-based quantum computer, instances of quantum types are stored on as many qubits as the bit-length of their classical counterparts.

For initialization, the \mintinline[bgcolor=bg]{text}{||} operator(s) create equal superposition of compile-time constant classical values, whereas \mintinline[bgcolor=bg]{text}{<qtype>.all()} creates equal superposition of all possible classical values (basis states) of a type. 
\vspace{4pt}
\begin{minted}[bgcolor=bg,breaklines,breakanywhere,frame=single,framesep=2mm]{text}
qbit b = 0 || 1;
qbit[] bArr = qbit.zeros(4);
qint i = 15; // basis state
qint j = 1 || 30 || 160;
qint k = qint.all(); // equal superposition of ALL int values
qfloat f = 2.5 || 3.5;
qcomplex z = (0.6 + 0.8i) || (sqrt(1/2) + sqrt(1/2)i);
qchar c = 'A' || 'B' || 'X' || 'Y';
qstring s = "Hello" || "World";
qbool tf = true || false;
qref r = &i || &j;
\end{minted}
\vspace{4pt}

Since all quantum values have an encoding in qubits, the compiler can apply conditional qubit-rotation techniques for state preparation to implement the \mintinline[bgcolor=bg]{text}{||} operator at the gate level~\cite{Mozafari21}, and Hadamard gates to implement the \mintinline[bgcolor=bg]{text}{all()} method.

\textbf{Measurement.} When assigning a quantum variable to a classical variable, the quantum variable is measured in the classical basis and collapses. The classical variable holds the result.

Below is an example with the following steps: (1) the program initializes the variables, 
(2) measures a \mintinline[bgcolor=bg]{text}{qref} which references two \mintinline[bgcolor=bg]{text}{qfloat} variables simultaneously,
(3) measures the \mintinline[bgcolor=bg]{text}{qfloat} corresponding to the resulting address from the previous step, which is stored in a \mintinline[bgcolor=bg]{text}{ref}, and (4) prints \mintinline[bgcolor=bg]{text}{float y}.

\vspace{4pt}
\begin{minted}[bgcolor=bg,breaklines,breakanywhere,frame=single,framesep=2mm]{text}
qfloat f = 2.5 || 3.5;
qfloat g = 3.14159 || 2.71828;
qref r = &f || &g;

ref x = r; // r collapses, x is either &f or &g
float y = *x; // *x is a qfloat, it collapses, y holds outcome

print(y);
\end{minted}
\vspace{4pt}

\section{Computation}\label{s:computation}
Quantum types are compatible with two forms of computation: classical computation (arithmetic, modifications...), and quantum computation.
Classical operations on quantum types are defined by their classical counterparts. This section explains how three basic common quantum-related operations are defined: phase shifts, interference, and conditional operations.

\textbf{Adding phase.} On every instance of a quantum type, programmers can call the \mintinline[bgcolor=bg]{text}{addPhase(shift,N)} method, where the first argument \mintinline[bgcolor=bg]{text}{shift} is a classical function that calculates a phase shift for each basis state and is expressed as an integer multiple of the angle \(\frac{2\pi}{N}\). 
The second argument, \mintinline[bgcolor=bg]{text}{N}, is a positive integer, a compile-time constant. 
The example code below brings \mintinline[bgcolor=bg]{text}{b} from \(\frac{1}{\sqrt{2}}|0\rangle+\frac{1}{\sqrt{2}}|1\rangle\) to \(\frac{1}{\sqrt{2}}|0\rangle-\frac{1}{\sqrt{2}}|1\rangle\).

The compiler can implement the \mintinline[bgcolor=bg]{text}{addPhase} method at the gate level by applying a generic phase kick-back technique \cite{Cleve98}. 
As for the classical code of \mintinline[bgcolor=bg]{text}{shift}, a compilation pass can convert it to an equivalent reversible quantum circuit \cite{Litteken20}.

\vspace{8pt}
\begin{minted}[bgcolor=bg,frame=single,framesep=2mm]{text}
qbit b = 0 || 1;
b.addPhase(shift, 2);

def shift(bit b) -> bit {
  if (b == 1) {
    return 1;
  } else {
    return 0;
  }
}
\end{minted}
\vspace{4pt}

\textbf{Creating interference.} On any instance of a quantum type, one can call \mintinline[bgcolor=bg]{text}{applyBipartiteInterference(split,pair)}, which\linebreak makes the classical basis states pairwise interfere with each other. 
The first argument, \mintinline[bgcolor=bg]{text}{split}, is a classical function that divides the set of all possible classical values (basis states) of the quantum type into two disjoint subsets, marked by the return value \mintinline[bgcolor=bg]{text}{true} or \mintinline[bgcolor=bg]{text}{false}. 
The second argument \mintinline[bgcolor=bg]{text}{pair} is a classical bijective map between the "true" and "false" subsets.\footnote{For convenience, it is also allowed that \mintinline[bgcolor=bg]{text}{pair} maps a value to itself, which causes the amplitude of the corresponding basis state to stay the same. 
If \mintinline[bgcolor=bg]{text}{split} and \mintinline[bgcolor=bg]{text}{pair} do not make up a one-to-one bipartition on the (sub)set of values not mapped to themselves, the behavior of \mintinline[bgcolor=bg]{text}{applyBipartiteInterference} is compiler-dependent.}

The interference caused is analogous to that of a Hadamard gate, \(H\), for qubits: if a classical basis state, \(|x\rangle\), in the "true" subset has amplitude \(a_x\), while its pair \(|y\rangle\) in the "false" subset has amplitude \(a_y\), then \(|x\rangle\) will end up with amplitude \(\frac{1}{\sqrt{2}}(a_x+a_y)\), while \(|y\rangle\) with amplitude \(\frac{1}{\sqrt{2}}(a_x-a_y)\).

In the example code below, \mintinline[bgcolor=bg]{text}{applyBipartiteInterference} brings \mintinline[bgcolor=bg]{text}{i} from \(a|80\rangle+b|81\rangle+c|144\rangle+d|145\rangle\) to \(\frac{1}{\sqrt{2}}(a+b)|80\rangle+\frac{1}{\sqrt{2}}(a-b)|81\rangle+\frac{1}{\sqrt{2}}(c+d)|144\rangle+\frac{1}{\sqrt{2}}(c-d)|145\rangle\), so \(|80\rangle\) interfered with \(|81\rangle\), and \(|144\rangle\) with \(|145\rangle\).

\vspace{8pt}
\begin{minted}[mathescape,bgcolor=bg,frame=single,framesep=2mm]{text}
qint i = 80 || 81 || 144 || 145;
...
// code that makes i = a|80>+b|81>+c|144>+d|145>
...
i.applyBipartiteInterference(split, pair);

def split(int i) -> bool {
  if (i % 2 == 0) {
    return true;
  } else {
    return false;
  }
}

def pair(int i) -> int {
  if (i % 2 == 0) {
    return i + 1;
  } else {
    return i - 1;
  }
}
\end{minted}
\vspace{4pt}

A more general interference, analogous to that of an arbitrary 1-qubit quantum gate, \(U=\begin{bmatrix}u_{11} & u_{12}\\ u_{21} & u_{22}\end{bmatrix}\), can be achieved by adding the elements of \(U\) as parameters (see code snippet below). 
In the previous example, this variant would have brought \mintinline[bgcolor=bg]{text}{i} to\linebreak \((u_{11}a+u_{12}b)|80\rangle+(u_{21}a+u_{22}b)|81\rangle+(u_{11}c+u_{12}d)|144\rangle+(u_{21}c+u_{22}d)|145\rangle\).\footnote{We note that the same interference can be reproduced (approximately, up to the precision allowed by the \mintinline[bgcolor=bg]{text}{int} data type) by using only $U=H$ and \mintinline[bgcolor=bg]{text}{addPhase} steps.}

\vspace{8pt}
\begin{minted}[bgcolor=bg,breaklines,breakanywhere,frame=single,framesep=2mm]{text}
i.applyBipartiteInterference(split, pair, u11, u12, u21, u22);
\end{minted}
\vspace{4pt}

The compiler can implement \mintinline[bgcolor=bg]{text}{applyBipartiteInterference} at the gate level by making use of the fact that \mintinline[bgcolor=bg]{text}{split} and \mintinline[bgcolor=bg]{text}{pair} create a one-to-one bipartition. This bipartition pairs basis states, which can be transformed to each other in a straightforward way, on classical "branches" of a superposition state.

\textbf{Conditional operation.} Programmers can apply a classical reversible operation on a classical value (basis state) of a quantum variable, conditioned on a (classical) logical expression evaluated on the classical values (in superposition) of another quantum variable. 
The resulting effect of the operation is the creation of entanglement between the two quantum variables.

The code snippet below produces entanglement between two \mintinline[bgcolor=bg]{text}{qchar} variables.
The \mintinline[bgcolor=bg]{text}{if (condition) {...}} block brings the composite quantum state of \mintinline[bgcolor=bg]{text}{c} and \mintinline[bgcolor=bg]{text}{t} from the product state\linebreak \(\left(\frac{1}{\sqrt{2}}|\text{'A'}\rangle+\frac{1}{\sqrt{2}}|\text{'B'}\rangle\right)\otimes|\text{'A'}\rangle\) to the entangled state \(\frac{1}{\sqrt{2}}|\text{'A'}\rangle|\text{'A'}\rangle+\frac{1}{\sqrt{2}}|\text{'B'}\rangle|\text{'B'}\rangle\).

\vspace{4pt}
\begin{minted}[bgcolor=bg,breaklines,breakanywhere,frame=single,framesep=2mm]{text}
qchar c = 'A' || 'B';
qchar t = 'A';

if (c == 'B') {
  t.increment(); // 'A' to 'B', t becomes entangled with c
}
\end{minted}
\vspace{4pt}

The compiler stores the result of the logical expression in an ancilla qubit. Conditioned on the ancilla state, the compiler first executes the body of the \mintinline[bgcolor=bg]{text}{if} statement, and then the uncomputation of the ancilla qubit.

\section{Evaluation}\label{s:eval}
\textbf{Universality.} The quantum data types are represented as qudits, typically representing quantum systems with more than two dimensions. 
For example, since \mintinline[bgcolor=bg]{text}{char} is encoded with 7-bit ASCII characters, a \mintinline[bgcolor=bg]{text}{qchar} is a qudit of dimension \(d=128\). 
Applying \mintinline[bgcolor=bg]{text}{addPhase} and \mintinline[bgcolor=bg]{text}{applyBipartiteInterference} can reproduce any single-qudit unitary transformation \cite{Luo2014}.\footnote{Up to the precision allowed by the numeric data types and the hardware.} 
Together with the \mintinline[bgcolor=bg]{text}{if} statement, which can create entanglement between single qudits, the programming language is universal with respect to any given quantum data type \cite{Brylinski2002}.

\textbf{Example: Grover algorithm with strings.} 
Similarly to classical programming, in certain situations it is useful to write a subroutine using low-level (assembly) instructions to achieve a highly optimized piece of code. To demonstrate this point, below is the code for a simple Grover search, where the goal is to find the string "ABC", marked by the \mintinline[bgcolor=bg]{text}{oracle(string s)} function.
\vspace{8pt}
\begin{minted}[bgcolor=bg,breaklines,breakanywhere,frame=single,framesep=2mm]{text}
qstring s = qstring.all();

int dim = qstring.dimension();
int numIter = ceil(pi / 4 * sqrt(dim));

// Grover iteration
for(int i = 0; i < numIter; i++) {
  s.applyOracle(oracle);
  s.invertAboutMean();
}

string res = s; // measure to get solution with high probability
print(res);

def oracle(string s) -> bool {
  if (s == "ABC") {
    return true;
  } else {
    return false;
  }
}
\end{minted}
\vspace{4pt}

One option to implement the \mintinline[bgcolor=bg]{text}{invertAboutMean()} method, which is available for any quantum type, is to convert the quantum variable to \mintinline[bgcolor=bg]{text}{qbit[]} and call a generic library method written for qubit arrays using quantum gates:
\vspace{8pt}
\begin{minted}[bgcolor=bg,frame=single,framesep=2mm]{text}
def invertAboutMean(qbit[] arr) {
  int L = arr.length;
  qbit[] anc = qbit.zeros(L);
  for(int i = 0; i < L; i++) {
    arr[i].H();
    arr[i].X();
  }
  anc[0].CNOT(arr[0]);
  for(int i = 1; i < L; i++) {
    anc[i].CCNOT(arr[i], anc[i-1]);
  }
  anc[L-1].Z();
  for(int i = L - 1; i > 0; i--) {
    anc[i].CCNOT(arr[i], anc[i-1]);
  }
  anc[0].CNOT(arr[0]);
  for(int i = 0; i < L; i++) {
    arr[i].X();
    arr[i].H();
  }
}
\end{minted}
\vspace{4pt}

Another possibility is not to expose the gate-level implementation at all, declaring the function as \mintinline[bgcolor=bg]{text}{native}, as shown below for the \mintinline[bgcolor=bg]{text}{qstring} type. 
In general, hiding the quantum gates forces the programmer to think about the impact of quantum operations using high-level classical data structures. 
For example, binary trees can be used to capture the overall impact of Hadamard gates applied in parallel at the qubit level.
\vspace{4pt}
\begin{minted}[bgcolor=bg,frame=single,framesep=2mm]{text}
def native invertAboutMean(qstring s);
\end{minted}
\vspace{4pt}
Finally, a third approach consists in applying \mintinline[bgcolor=bg]{text}{addPhase},\linebreak \mintinline[bgcolor=bg]{text}{applyBipartiteInterference} and \mintinline[bgcolor=bg]{text}{if} statements in the body of \mintinline[bgcolor=bg]{text}{invertAboutMean()}, instead of low-level gates. This way, each quantum type would need a bespoke implementation.

\section{Related work}\label{s:rw}
In today's commonly used quantum programming languages, high-level abstraction features help developers express quantum algorithms operating on qubits using quantum gates. 
Such features include classical control flow, gate modifiers to automatically generate inverse and controlled versions of unitary subroutines, as well as automatic uncomputation facilities~\cite{Bichsel20,Cross18,Cross22,Seidel23,Svore18}. 
Other approaches include synthesizing large quantum circuits from high-level (template) functional models~\cite{Classiq} or the parsing of classical code to automatically detect quantum speedups and replace the relevant parts with quantum circuits~\cite{Horizon}.

Regarding data types, several quantum programming languages introduce the concept of "quantum integers". 
In Silq~\cite{Bichsel20}, a quantum integer is an array of qubits and it is manipulated directly at the qubit level.
In Quipper~\cite{Green13}, an arithmetic library defines a fixed-size quantum integer type called QDInt, that supports classical arithmetic operations, but without QDInt-level operations to holistically manipulate QDInt instances in a universal manner. 
There is also a similar library that introduces the FPReal type. There are currently no genuine quantum data types other than qubits and arrays of qubits, and programming mostly consists of gate-level manipulations.

\section{Conclusions}\label{s:conc}
The lack of high-level abstractions for quantum programs is one of the major issues for software developers who want to write quantum code.
It hinders developers' ability to produce new quantum algorithms.
%Despite learning the intricacies of qubits and quantum gates, developers generally have difficulties using such structures with higher-level abstractions.
%As an example, while there exists a quantum search algorithm, it is very difficult to code real search algorithms if strings or other complex types do not exist.

This article proposes Rhyme, a quantum programming language that includes high-level quantum types based on familiar classical numeric and text data types. 
In that language, the amplitudes of quantum superposition states can be manipulated by writing solely classical functions.
We presented results in the form of code snippets.
We plan to develop a full-fledged programming language based on the ideas presented here, hosted on \url{https://quantumpl.org}. 

%\bibliographystyle{ACM-Reference-Format}
%\bibliography{biblio}
\input sample-manuscript.bbl

\end{document}

%% file: sample-manuscript.bbl
%%% -*-BibTeX-*-
%%% Do NOT edit. File created by BibTeX with style
%%% ACM-Reference-Format-Journals [18-Jan-2012].